# Capturing the Production of Innovative Ideas:
# An Online Social Network Experiment and "Idea Geography" Visualization

Yiding Cao[*], Yingjun Dong, Minjun Kim, Neil G. MacLaren, Ankita Kulkarni,
Shelley D. Dionne, Francis J. Yammarino, & Hiroki Sayama[*]
Binghamton University, State University of New York, Binghamton, NY13902, USA
[*]ycao20@binghamton.edu

**Abstract**

Collective design and innovation are crucial in organizations. To investigate how the collective design and innovation processes would be affected by the diversity of knowledge and background of collective individual members, we conducted three collaborative design task experiments which involved nearly 300 participants who worked together anonymously in a social network structure using a custom-made computer-mediated collaboration platform. We compared the idea generation activity among three different background distribution conditions (clustered, random, and dispersed) with the help of the "doc2vec" text representation machine learning algorithm. We also developed a new method called "Idea Geography" to visualize the idea utility terrain on a 2D problem domain. The results showed that groups with random background allocation tended to produce the best design idea with highest utility values. It was also suggested that the diversity of participants' backgrounds distribution on the network might interact with each other to affect the diversity of ideas generated. The proposed idea geography successfully visualized that the collective design processes did find the high utility area through exploration and exploitation in collaborative work.

## 1. Introduction

Collective design and innovation have been an important research subject in social sciences and engineering, because its processes are needed for successful solution development for many real-world problems in an organization (Kerr & Tindale, 2004; Lu, ElMaraghy, Schuh & Wilhelm, 2007). It is therefore essential to consider how to improve the quality and efficiency of the collective design and innovation processes. Large-scale design processes that occur in the collective design and innovation involve interaction and interdependence among multiple individuals with task-related diversity (Sayama & Dionne, 2015; Dionne, Sayama & Yammarino, 2019). In particular, the interdisciplinary background within teams has been shown to be positively correlated with both quantitative and qualitative task performance (Horwitz & Horwitz, 2007; Koh, 2008; Salas, Rosen & DiazGranados, 2010). This suggests that investigating the effects of expertise on collective performance would be a promising area of further research to improve the quality of collective design and innovation.

The organizational structure under which the collective design proceeds is usually complex, which makes it harder to investigate the collective dynamic performance in realistic organizational settings (Braha & Bar-Yam, 2004, 2007; Kijkuit & Van den Ende, 2010; McCubbins, Paturi & Weller, 2009). Previous studies on this problem were limited in several ways: (1) The network size was significantly smaller than that of most real-world collective design cases, (2) the duration of the collective task was significantly shorter than that of most real-world collective design cases, and (3) the tasks used in models or experiments were quite simple and not open-ended (Becker, Woolley, Chabris & Pentland, 2010; Mason & Watts, 2012; Brackbill & Centola, 2017).

The objective of the present study is to experimentally investigate how the diversity of background of individual members will affect the effectiveness of design and innovation processes at collective levels. Our approach involves a combination of theoretical agent-based simulation models (Sayama & Dionne, 2015; Dionne, Sayama, Hao & Bush, 2010; Dionne, Sayama & Yammarino 2019) and online human-subject social network experiments. This paper will describe our online social network experiments and report some initial findings.

Previous human-subject studies typically evaluated team performance by basic variables such as the number of ideas generated, average scores, and win rates (Sayama & Dionne, 2015; Sapienza, Zeng, Bessi, Lerman & Ferrara, 2018). In contrast, the present study not only measured such basic variables, like the number of posted ideas and the utility score of final ideas, but also measured quantitative similarity between produced ideas using "doc2vec", a text embedding machine-learning algorithm (Le & Mikolov, 2014). We used the doc2vec algorithm to convert text-format design ideas to numerical vectors and performed further quantitative analyses. We also developed a new method called "Idea Geography" to visualize the idea utility terrain on a 2D problem domain. The utility terrain visualized with Idea Geography can help determine which region(s) in the problem space would be more (or less) promising for further exploration.

The rest of the paper is structured as follows. Section 2 introduces details of three online human-subject social network experiments we have conducted, including experimental procedures, data collection, and task descriptions for the three experimental sessions. Section 3 describes data analysis processes and methods, including the "Idea Geography" method we developed. Section 4 presents the results. Section 5 discusses the findings and concludes the paper with future directions.

**2. Online social network experiment**

We designed and conducted three online experiments using a custom-made web-based computer-mediated collaboration (CMC) platform with an interface is similar to Twitter. This platform was implemented using Python and Flask. This platform allows participants to submit ideas in response to the assigned design task, see other participants' ideas, and add comments and "likes" to those ideas. We recruited a multidisciplinary group of students at a mid-size US public university in the Fall 2018 and Spring 2019 semesters to participate in the experiments.

Participants were undergraduate/graduate students majoring in Engineering or Management. They were allowed to sign up for one experimental session (i.e, they were not allowed to take part in this experiment more than once). Each experimental session involved 64~77 participants who worked on an open-ended collective design task for two weeks.

## 2.1. Experimental procedure

To participate in an experimental session, participants were required to fill out an experimental registration form to provide their academic major and a written description of why they selected their major, as well as their academic knowledge, technical skills, career interest areas, hobbies or extracurricular activities, and/or any other information related to their background (this information is called simply "background" in this paper). This narrative information was converted into a numerical vector using the doc2vec algorithm. The academic major and vectors representing the background characteristics of the participants were used to allocate the participants into the following three groups: (1) spatially clustered, i.e., participants with similar background were placed together as social neighbors; (2) randomly distributed, i.e., participants were randomly placed regardless of their background; and (3) dispersed, i.e., participants with different background were placed together as social neighbors. These groups were configured to be similar to each other in terms of the amount of within-group background variations; they differed only with regard to spatial distributions of background variations. The underlying social network topology was a spatially clustered regular network made of 21~26 members with degree four, in all of the three groups. Participants could observe only their immediate neighbor's activities and would not directly see activities of other nonadjacent participants.

At the beginning of each experimental session, participants were provided an overall objective of the collective design task and instructions of how to use the experimental platform. This online experimental session lasted for two weeks, during which participants were requested to log in to the experimental platform using anonymized usernames and spend at least 15 minutes each weekday, working on the assigned collective design task with their collaborators (i.e., anonymous neighbors in the social network). Their participation and actions were logged electronically in the server and monitored by the experimenters on a regular basis.

The collective design task description was displayed at the top of the experimental platform interface (Figure 1). This interface allows participants to generate and post new ideas using the input box below the task description, read their collaborators' ideas in the timeline shown below the input box, and like and comment on others' ideas as well, like in Twitter. On each weekday during the experimental session, participants were requested to post ideas on the platform, discuss the task by reading, commenting, and liking their collaborators' ideas. By potentially utilizing their collaborators' ideas and comments, participants were expected to continuously elaborate and improve their idea quality over time on the platform.

After the two-week experimental session was over, each participant was asked to submit an end-of-session survey form to provide three final ideas they chose for the assigned design task. These final ideas were later evaluated by third-party experts who were not involved in the experiments. These evaluation results were used to quantitatively assess the utility values of the final ideas made by each group. The participants were also asked in the survey form to answer questions about their overall experience, level of knowledge and understanding about their organizational neighbors, self-evaluation of own contribution to the collaborative process, and personal evaluation of the final designs.

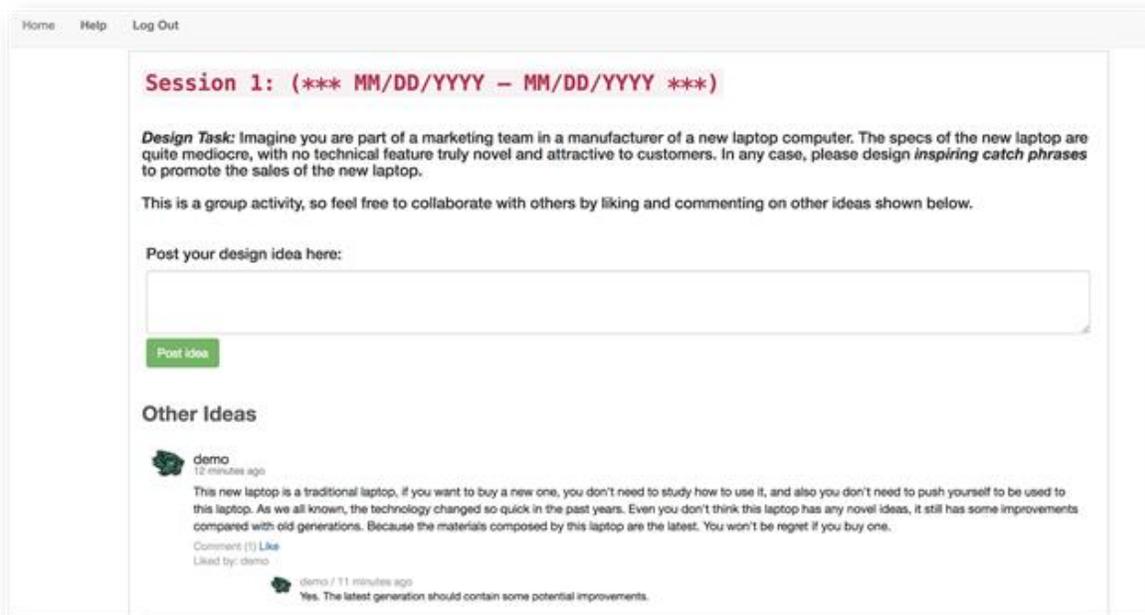

Figure 1: Screen shot of the experimental platform

### 2.2. Collective design tasks

There were two different design tasks used for the experimental sessions. The tasks were open-ended textual design tasks with no obvious solutions immediately available to anyone.

The task for experiment session I was to create slogans, taglines or catch phrases for marketing a laptop. This task worked successfully for students with diverse backgrounds in previous work (Sayama & Dionne, 2015). This task was used in two experimental sessions conducted in two separate semesters (Fall 2018 and Spring 2019). In the Fall 2018 session I, participants were a relatively balanced mixture of Engineering and Management majors. A total of 64 students participated in this session. In the Spring 2019 session, participants were predominantly majoring in Management. A total of 66 students participated in this session. The final designs submitted by the participants who worked on this task were evaluated on a 5-point Likert scale by Marketing PhD students who did not participate in the experiment.

The task for experiment session II was to write a story or a complete fiction within a word count limit. This experimental session was conducted once in Spring 2019. A total of 74 students participated in this session who were predominantly majoring in Management. The final designs submitted by the participants who worked on this task were evaluated on a 5-point Likert scale by professional staff members on campus who had educational background and professional experience and expertise in creative writing and communication.

## 3. Data analysis methods

We applied a series of quantitative data analyses to investigate the effects of background distribution on the group performance. The participants' activity records were utilized as the dataset for these analyses. First, the numbers of daily posts and submitted final ideas were measured as a characteristic of each group. We compared both time-series and distribution of numbers of daily posts among the three groups for each experimental session. We also measured the Euclidean distances between the posted ideas converted into numerical vectors using doc2vec, and then created a distance matrix of the ideas. Furthermore, the ideas were visualized as points in a 2-dimensional problem space using the dimension-reduced idea vectors, which provided the basis to construct the "Idea Geography". Figure 2 shows an overview of our data analysis methods.

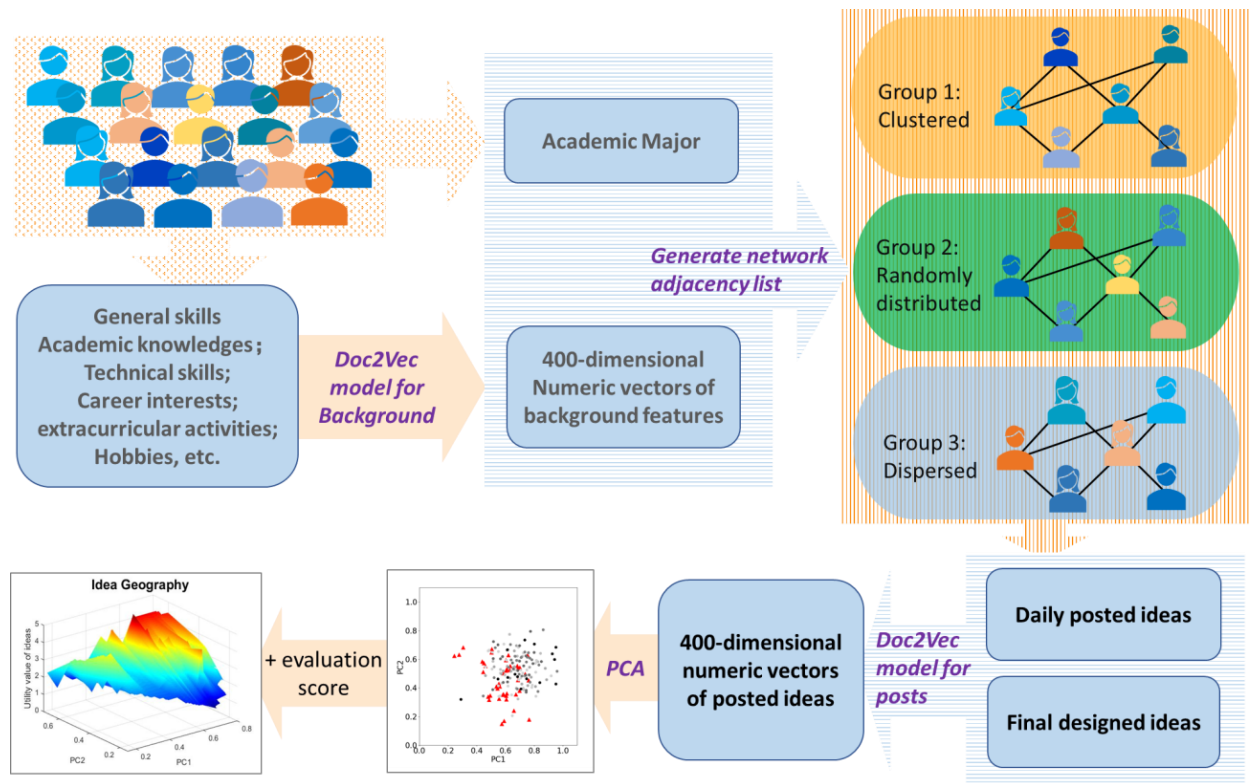

Figure 2: An overview of data analysis methods

### 3.1. Doc2Vec

The data acquired from the registration forms, experimental records, and the end-of-session survey forms were mostly in plain text format, which would be hard to analyze using traditional quantitative analysis methods. Therefore, it was necessary to convert the text data to numerical data using text representation algorithms. There are many text embedding algorithms, such as Bag of Words (BOW) (Harris, 1954), Latent Dirichlet Allocation (LDA) (Blei, Ng & Jordan, 2003), word2vec (Mikolov, Sutskever, Chen, Corrado & Dean, 2013), and doc2vec (Le & Mikolov, 2014) which was used in this study. Doc2vec, an adaptation of word2vec, is an unsupervised machine learning algorithm that can generate numerical vectors as a representation of sentences, paragraphs, or documents. Compared to other algorithms, doc2vec can provide a better text representation with a lower prediction error rate, because it can recognize the word ordering and semantics of words which are not accounted by other algorithms (Lau & Baldwin, 2016).

In this study, the whole set of written descriptions of background submitted by participants in the registration forms for each experimental session was used to build a doc2vec model of the participants' background for that session. The outputs of this doc2vec model were given in the form of 400-dimensional numerical vectors, which were combined with the self-reported academic major information to quantitatively represent the background features of the participants.

The daily posts and final ideas were also converted to 400-dimensional numerical vectors using doc2vec. The doc2vec model for the ideas posted during experiment session I was generated using the combined set of all the ideas obtained from both Fall 2018 and Spring 2019 session I. The doc2vec models for the ideas posted during experiment session II were generated using the sets of ideas obtained from the Spring 2019 session II.

### 3.2 Principal component analysis

Many of the 400 dimensions in the vectors obtained with doc2vec were undoubtedly correlated with each other, making the dataset highly redundant. Principal component analysis (PCA) was therefore applied to the set of idea vectors for each experimental session to reduce dimensionality and visualize the idea distribution in a 2D space using the first two principal components (PC1, PC2). The 2D principal component space offered an efficient way to monitor the locations of ideas and also provided the basis to construct the "Idea Geography" method explained in the following section.

### 3.3 Idea Geography

The "Idea Geography" visualization method was developed specifically for this study. The average evaluation scores of the final ideas were used as the elevations at the 2D idea points to construct a utility terrain for each experiment session in the 2D principal component space, which we call *idea geography*. From the landscape of idea geography, we can find mountain areas which represent regions in the problem space populated by ideas with high evaluation scores. We can also find valleys where the ideas would have low evaluation scores. In terms of physical geography and environmental studies, the terrain structure of a region is important for determining its suitability for human settlement (Olwig, 1996), water flow patterns (Baker & Capel, 2011), and other properties of the region. In similar ways, with idea geography, the terrain of a problem space can intuitively reveal which areas in the problem space would have high or low utilities and would be suitable for further exploration. The visualization using idea geography may also help a leader or an organization manager to monitor the status of collaborative activities of a group.

## 4 Results

### 4.1 Experiment session I: Catch phrase design

The end-of-session survey forms of Fall 2018 session I and Spring 2019 session I revealed that 89% and 75% of the participants, respectively, stated that they had a good overall experience in the experiment.

Figure 3 compares the numbers of daily ideas and final designs among three groups for the Fall 2018 and Spring 2019 session I. In both sessions, there was no statistically significant difference regarding the number of ideas among the three groups.

Figure 4 compares the average distance of ideas of each working day using the 400-dimensional idea vectors. In the Fall 2018 session I, a statistically significant difference was detected in terms of the average distance of ideas between Group 1 and Group 2 ($p = 0.012$), and Group 1 and Group 3 ($p = 0.0091$). In the Spring 2019 session I, there was no statistically significant difference regarding the average distance of ideas among the three groups.

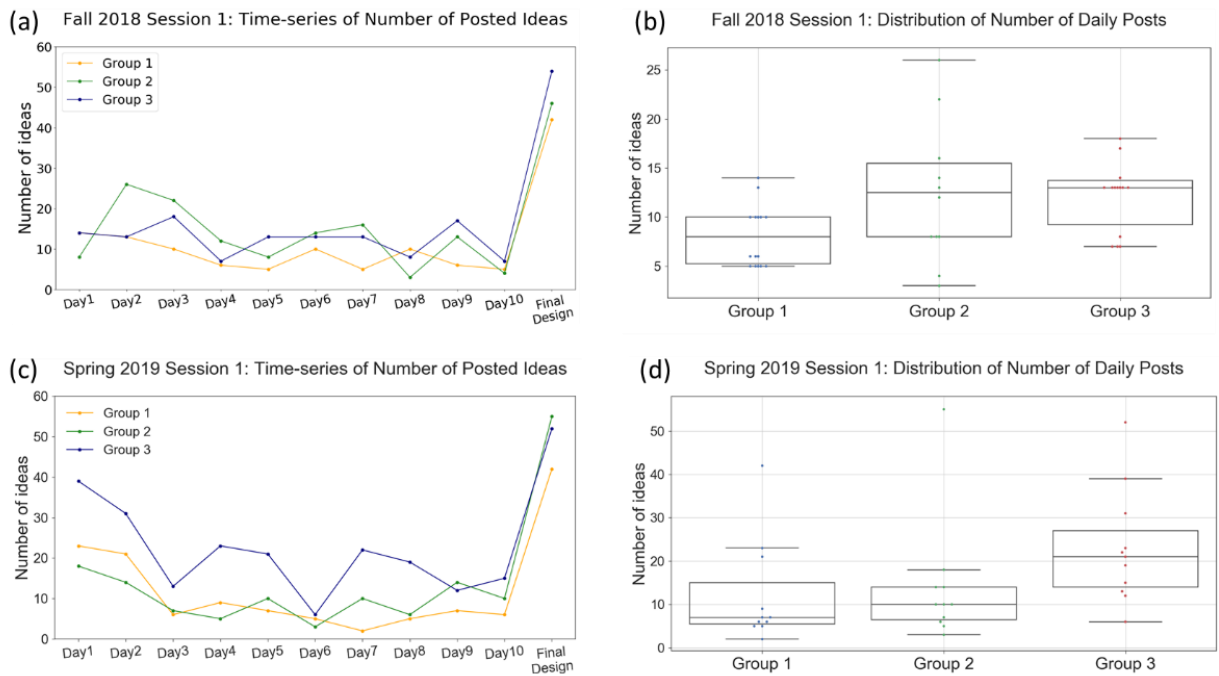

Figure 3: Fall 2018 and Spring 2019 session I: Number of daily posts and final ideas

The daily ideas and final designs of three groups were visualized as idea points in the 2D principal component problem space for two experimental sessions (Figure 5). Group 1 with clustered background allocation produced a broader daily idea distribution but a more concentrated final design distribution than the other two groups. In Spring 2019 session I, three groups all have concentrated daily idea and final design distributions.

Figure 6 shows the idea geography visualizations of the Fall 2018 and Spring 2019 session I. There is a clearly identifiable utility mountain area in each idea geography, where most of the submitted final designs were concentrated. In both sessions, Group 2 with random background allocation produced the best final design with highest utility value (Fall 2018: 4.75; Spring 2019: 4.667).

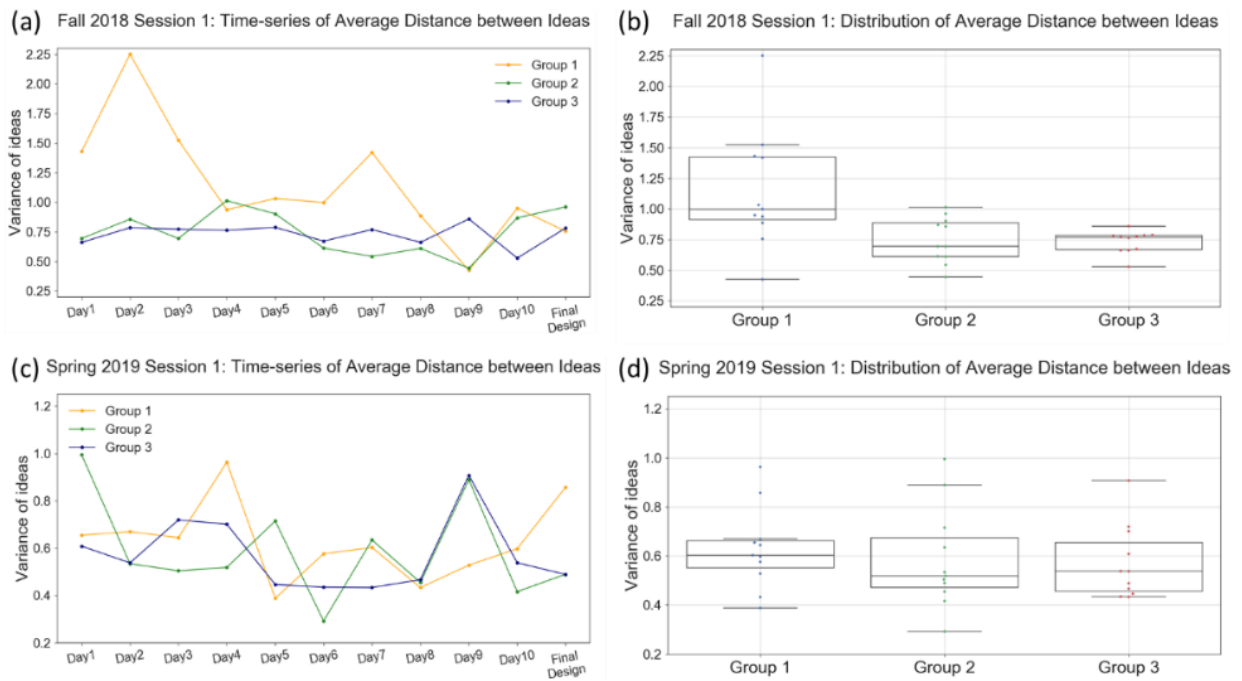

Figure 4: Fall 2018 and Spring 2019 session I: Average distance of ideas

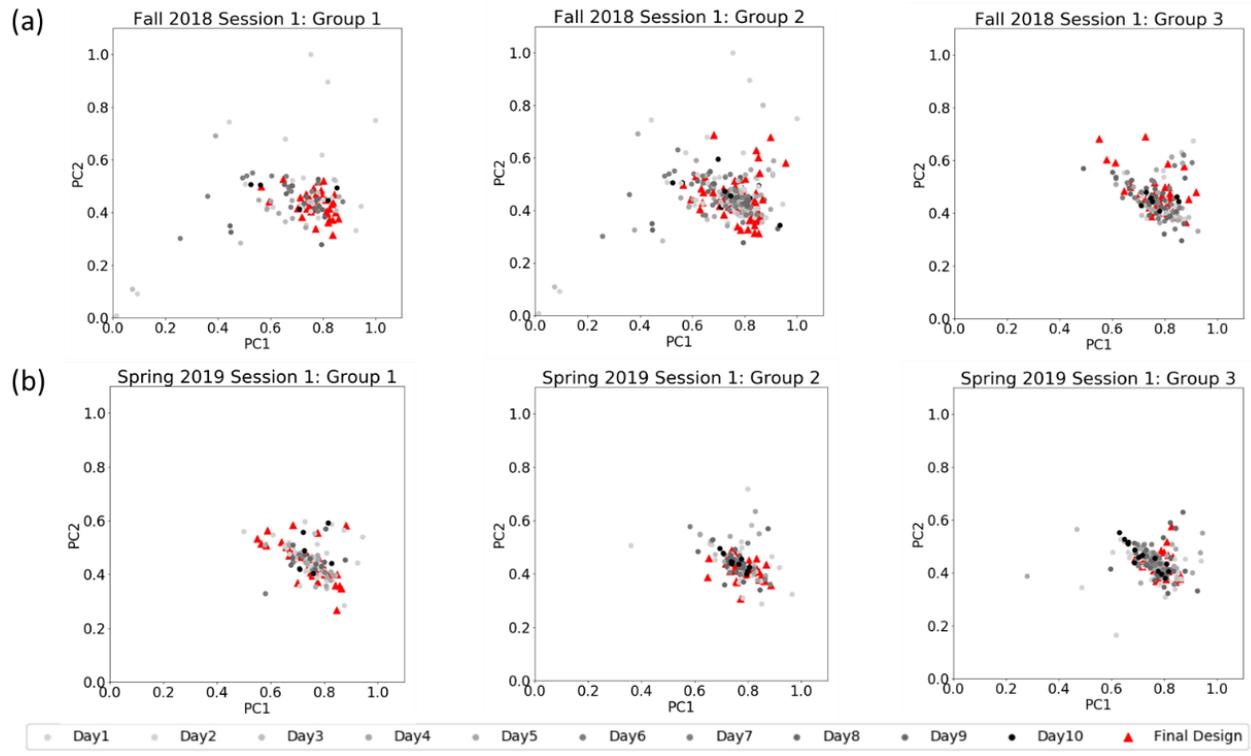

Figure 5: Fall 2018 and Spring 2019 session I: Distribution of idea points in 2D problem space. Note: Daily ideas from Day1 to Day10 are marked as circles with color ranging from light grey to black; Final designs are marked as red triangles.

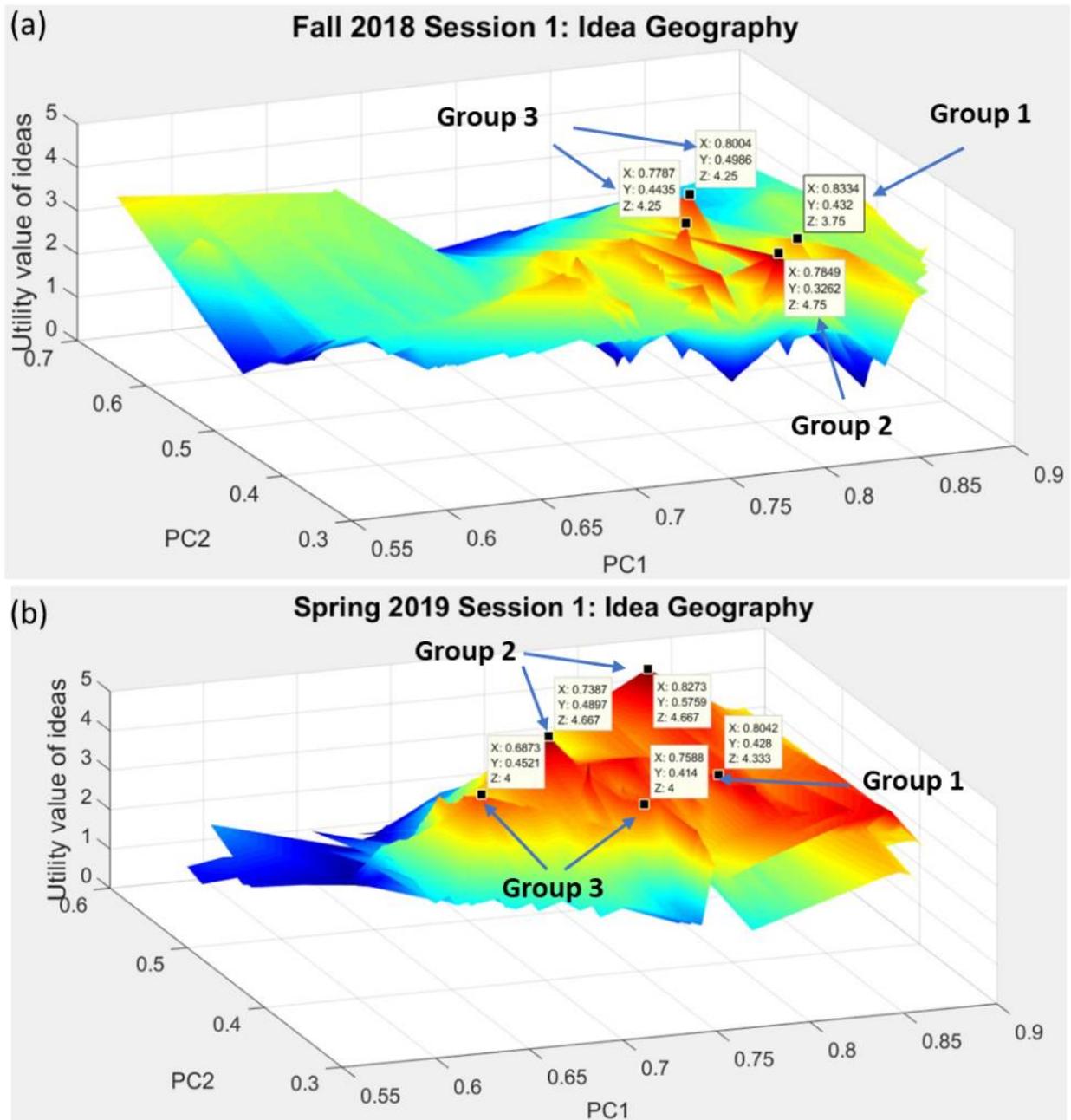

Figure 6: Fall 2018 and Spring 2019 session I: Idea Geography

## 4.2 Experiment session II: Story design

In the end-of-session survey form of this experimental session, 71% of the participants stated that they had a good overall experience in the experiment.

Figure 7 compares the numbers of daily ideas and final designs among three groups. There was no statistically significant difference regarding the number of ideas among the three groups in this session.

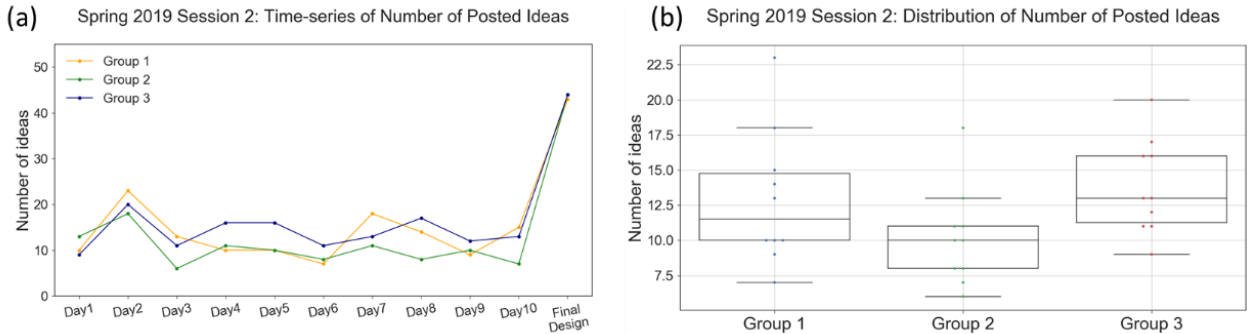

Figure 7: Spring 2019 session II: Number of daily posts and final ideas

Figure 8 compares the average distance of ideas of each working day in the Spring 2019 session II. There was no statistically significant difference detected regarding the average distance either.

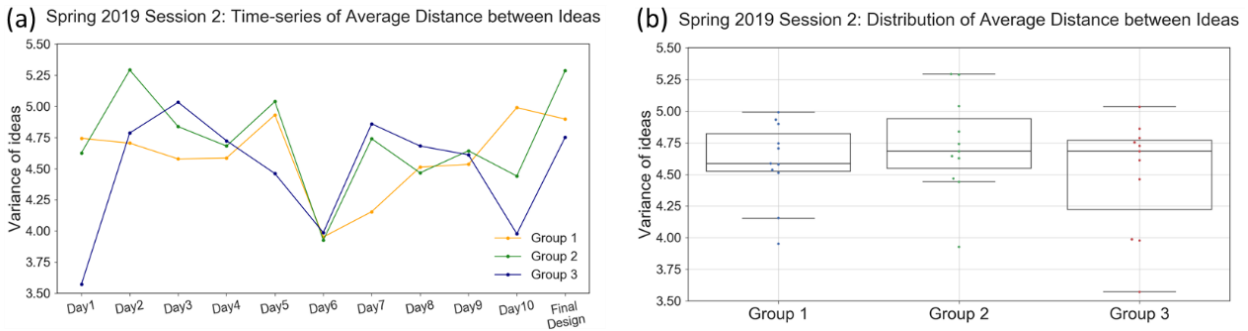

Figure 8: Spring 2019 session II: Average distance of ideas

The daily ideas and final designs generated by the three groups in the Spring 2019 session II were visualized in the 2D principal component problem space (Figure 9). The three groups showed similar patterns of both the daily idea and final design distribution in the problem space.

Figure 10 shows the idea geography visualization of the Spring 2019 session II. We can see again that there is a clearly identifiable utility mountain area, where most of the submitted final designs were concentrated. Interestingly, Group 2 with random background allocation again produced the best final design with highest utility value (5.0).

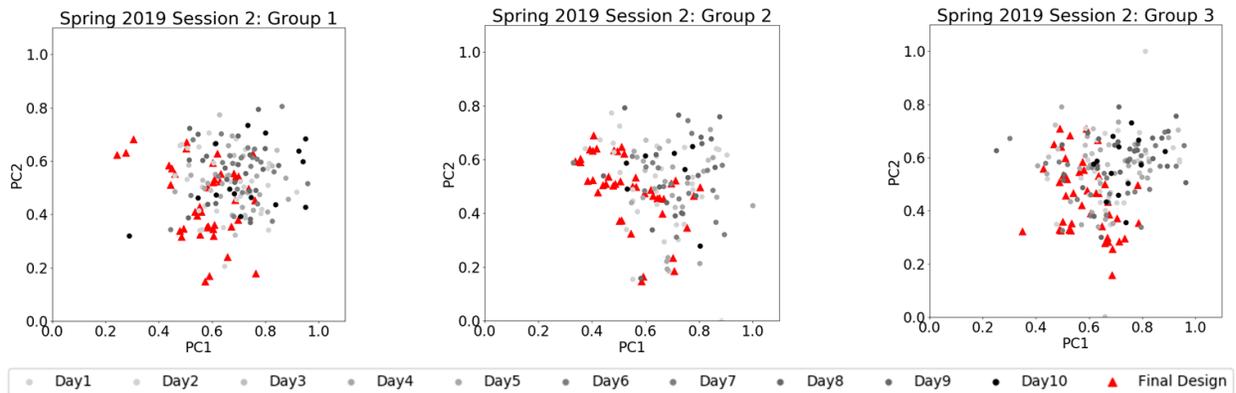

Figure 9: Spring 2019 session II: Distribution of idea points in 2D problem space. Note: Daily ideas from Day1 to Day10 are marked as circles with color ranging from light grey to black; Final designs are marked as red triangles.

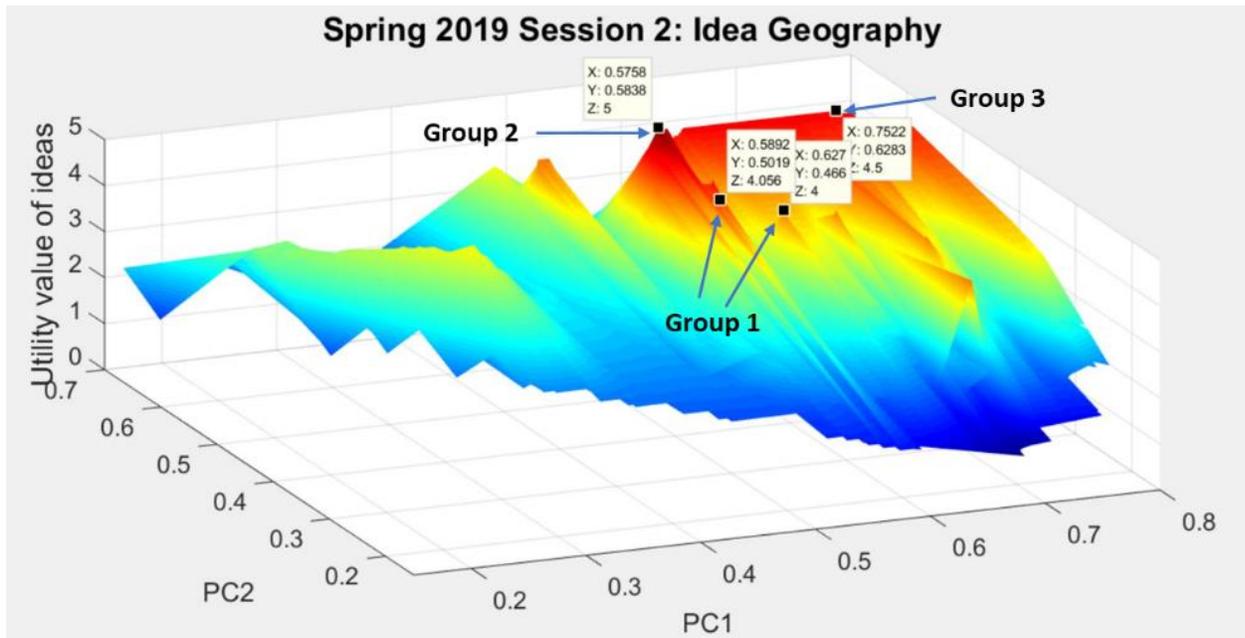

Figure 10: Spring 2019 session II: Idea Geography

## 5 Discussions

Our results showed no statistically significant difference in the numbers of daily ideas among three groups in any of the three experimental sessions. This may imply that the activity level of individual participants was not affected by the allocation of background in the social network.

Meanwhile, a clear statistical significance was detected in terms of the average distance of ideas for the Fall 2018 session I, in which Group 1 (with clustered background allocation) produced more diverse ideas than the other two groups. However, this was not observed in the Spring 2019 session I. The key difference between Fall 2018 and Spring 2019 was in the variety of participants' backgrounds, i.e., Fall 2018 had a balanced mixture of Engineering and Management majors, while Spring 2019 was predominantly Management majors. We hypothesize that, when the participants' backgrounds were diverse and they were spatially clustered based on their similarity (this occurred only in Group 1 of the Fall 2018 Session I in our study), different parts of the network would explore the problem space in different directions without much mixture, and therefore, the average distance between generated ideas would naturally go larger. This also explains why the same phenomenon was not observed in Group 2 and 3 in Fall 2018 or in any groups/sessions in Spring 2019.

The most interesting finding obtained so far was that, in all the experimental sessions for which idea geography was generated, Group 2 with random background allocation produced the best final design with highest utility value. This seemingly puzzling observation may be explained by considering how much background diversity each participant was exposed to locally. Namely, in either Group 1 or Group 3, each participant would be connected to their neighbors that were relatively homogeneous background-wise (i.e., the neighbors should be similar to the focal participant in Group 1, while they should be the opposite of the focal participant in Group 3). Group 2 with random background allocation should have realized the most diverse local neighbors around each participant, which, we hypothesize, may have contributed to the enhancement of innovation search processes in Group 2.

Our experiments are still ongoing to collect more data. With additional experimental data, we hope to test these findings and hypotheses described above.

## 6 Conclusions

In this paper, we conducted a series of online human-subject experiments to examine and monitor the collective design and innovation processes on three different open-ended text design tasks. The doc2vec algorithm was used to quantify text-based information, including participants' backgrounds and ideas generated, which allowed for numerical characterization and control of similarities and differences between different background or ideas. The results we obtained so far indicated that, when participants with various background were randomly placed on the network, the group tended to find the best design ideas. The results also indicated potential interaction between background distribution and the diversity of participants' backgrounds. More data are needed to confirm these observations and test our hypotheses to explain them.

This study also proposed the "Idea Geography" method, which was successfully demonstrated as an effective way to visualize the behavior and performance of the group's collaborative work. The fact that most of the submitted final designs clustered near the mountain areas in the idea geography indicates that the collective innovation and design processes were able to find the high utility regions through exploration and exploitation.

**Acknowledgment**

This material is based upon work supported by the National Science Foundation under Grant #1734147.

**References**

Baker, N.T., & Capel, P.D., (2011). "Environmental factors that influence the location of crop agriculture in the conterminous United States": *U.S. Geological Survey Scientific Investigations Report, 2011–5108*

Becker, J., Brackbill, D., & Centola, D. (2017). Network dynamics of social influence in the wisdom of crowds. *Proceedings of the National Academy of Sciences, 114(26), E5070-E5076.*

Blei, D, Ng, A., and Jordan, M. (2003). Latent dirichlet allocation. *Journal of Machine Learning Research*, 3:993–1022

Braha, D., & Bar-Yam, Y. (2004). Topology of large-scale engineering problem-solving networks. *Physical Review E, 69(1), 016113.*

Dionne, S. D., Sayama, H., Hao, C., & Bush, B. J. (2010). The role of leadership in shared mental model convergence and team performance improvement: An agent-based computational model. *Leadership Quarterly, 21(6), 1035-1049.*

Dionne, S. D., Sayama, H., & Yammarino, F. J. (2019). Diversity and social network structure in collective decision making: Evolutionary perspectives with agent-based simulations. *Complexity, 2019, 7591072.*

Harris, Z. (1954) Distributional Structure, *WORD, 10:2-3, 146-162*

Horwitz, S. & Horwitz, I. (2007). The effects of team diversity on team outcomes: A meta-analytic review of team demography. *Journal of Management, 33, 987-1015.*

Kerr, N. L., & Tindale, R. S. (2004). Group performance and decision making. *Annual Review of Psychology, 55, 623-655.*

Koh, W. T. H. (2008). Heterogeneous expertise and collective decision-making. *Social Choice and Welfare, 30(3), 457-473*

Kijkuit, B., & van den Ende, J. (2010). With a little help from our colleagues: A longitudinal study of social networks for innovation. *Organization Studies, 31(4), 451-479.*

Lau, J.H., & Baldwin.,T., (2016). An empirical evaluation of doc2vec with practical insights into document embedding generation. *Proceedings of the 1st Workshop on Representation Learning for NLP (2016), 78-86, Berlin, Germany.*

Le, Q., and Mikolov, T. (2014). Distributed representations of sentences and documents. *In Proceedings of the 31st International Conference on Machine Learning (ICML 2014), p1188–1196, Beijing, China.*

Lu, S. Y., ElMaraghy, W., Schuh, G., & Wilhelm, R. (2007). A scientific foundation of collaborative engineering. *CIRP Annals-Manufacturing Technology, 56(2), 605-634.*

Mason, W., & Watts, D. J. (2012). Collaborative learning in networks. *Proceedings of the National Academy of Sciences, 109(3), 764-769.*

McCubbins, M. D., Paturi, R., & Weller, N. (2009). Connected coordination network structure and group coordination. *American Politics Research, 37(5), 899-920.*

Mikolov, T., Sutskever, I., Chen, K., Corrado,G., & Dean, J. (2013). Distributed representations of phrases and their compositionality. In *Advances on Neural Information Processing Systems.* arXiv:1310.4546v1

Olwig K.R. (1996), Recovering the Substantive Nature of Landscape, *Annals of the A.A.G (1996), 86(4), 630-653*

Salas, E., Rosen, M.A., & DiazGranados, D. (2010). Expertise-Based intuition and decision making in organizations. *Journal of Management, 36(4), 941-973*

Sapienza, A., Zeng, Y., Bessi, A., Lerman, K., & Ferrara, E. (2018). Individual performance in team-based online games. *Royal society open science, 5 (6), 180329*

Sayama, H., & Dionne, S. D. (2015). Studying collective human decision making and creativity with evolutionary computation. *Artificial Life, 21(3), 379-393.*

Woolley, A.W., Chabris, C.F., & Pentland, A. (2010). Evidence for a collective intelligence factor in the performance of human groups. *Science 330: 686–88*